\documentclass[12pt]{article}
\catcode`\@=11
\usepackage{amsmath}
\usepackage{amssymb}
\newcommand{\be}{\begin{equation}}
\newcommand{\ee}{\end{equation}}
\newcommand{\ba}{\begin{eqnarray}}
\newcommand{\ea}{\end{eqnarray}}
\newcommand{\bas}{\begin{eqnarray*}}
\newcommand{\eas}{\end{eqnarray*}}
\def\12{\frac{1}{2}}
\def\fr{\frac}
\def\pr{\partial}

\def\ww{\wedge}

\def\ra{\rightarrow}

\def\a{\alpha}
\def\b{\beta}
\def\g{\gamma}
\def\G{\Gamma}

\def\d{\delta}
\def\e{\epsilon}

\def\h{\eta}

\def\l{\lambda}
\def\L{\Lambda}
\def\m{\mu}
\def\n{\nu}

\def\9{\rho}
\def\s{\sigma}
\def\t{\tau}

\def\R{{\mathbb{R}}}
\def\ti{\tilde}

\def\@normalsize{\@setsize\normalsize{15pt}\xiipt\@xiipt
\abovedisplayskip 14pt plus3pt minus3pt%
\belowdisplayskip \abovedisplayskip
\abovedisplayshortskip  \z@ plus3pt%
\belowdisplayshortskip  7pt plus3.5pt minus0pt}
\def\small{\@setsize\small{13.6pt}\xipt\@xipt
\abovedisplayskip 13pt plus3pt minus3pt%
\belowdisplayskip \abovedisplayskip
\abovedisplayshortskip  \z@ plus3pt%
\belowdisplayshortskip  7pt plus3.5pt minus0pt
\def\@listi{\parsep 4.5pt plus 2pt minus 1pt
            \itemsep \parsep
            \topsep 9pt plus 3pt minus 3pt}}

\def\underline#1{\relax\ifmmode\@@underline#1\else
        $\@@underline{\hbox{#1}}$\relax\fi}
\@twosidetrue
\relax

\catcode`@=12

\catcode`\@=11

\def\section{\@startsection{section}{1}{\z@}{3.5ex plus 1ex minus
   .2ex}{2.3ex plus .2ex}{\large\bf}}

\def\ps@headings{\def\@oddfoot{}\def\@evenfoot{}
\def\@oddhead{\hbox{}\hfill
        \makebox[.5\textwidth]{\raggedright\ignorespaces --\thepage{}--
        \hfill }}
\def\@evenhead{\@oddhead}
\def\subsectionmark##1{\markboth{##1}{}} }
\renewcommand{\subsection}[1]{\addtocounter{subsection}{1}
\vspace{2.5mm}\par\noindent {\em \thesubsection . #1}\par
 \vspace{0.5mm} }

\ps@headings

\catcode`\@=12

\relax

\def\figcap{\section*{Figure Captions\markboth
        {FIGURECAPTIONS}{FIGURECAPTIONS}}\list
        {Fig. \arabic{enumi}:\hfill}{\settowidth\labelwidth{Fig. 999:}
        \leftmargin\labelwidth
        \advance\leftmargin\labelsep\usecounter{enumi}}}
 \relax
\def\tablecap{\section*{Table Captions\markboth
        {TABLECAPTIONS}{TABLECAPTIONS}}\list
        {Table \arabic{enumi}:\hfill}{\settowidth\labelwidth{Table
999:}
        \leftmargin\labelwidth
        \advance\leftmargin\labelsep\usecounter{enumi}}}
 \relax
\def\reflist{\section*{References\markboth
        {REFLIST}{REFLIST}}\list
        {[\arabic{enumi}]\hfill}{\settowidth\labelwidth{[999]}
        \leftmargin\labelwidth
        \advance\leftmargin\labelsep\usecounter{enumi}}}
 \relax

\catcode`\@=11

\def\marginnote#1{}
\newcount\hour
\newcount\minute
\newtoks\amorpm
\hour=\time\divide\hour by60
\minute=\time{\multiply\hour by60 \global\advance\minute by-
\hour}
\edef\standardtime{{\ifnum\hour<12 \global\amorpm={am}%
    \else\global\amorpm={pm}\advance\hour by-12 \fi
    \ifnum\hour=0 \hour=12 \fi
    \number\hour:\ifnum\minute<100\fi\number\minute\the\amorpm}}
\edef\militarytime{\number\hour:\ifnum\minute<100\fi\number\minute}
\def\draftlabel#1{{\@bsphack\if@filesw {\let\thepage\relax
  \xdef\@gtempa{\write\@auxout{\string
    \newlabel{#1}{{\@currentlabel}{\thepage}}}}}\@gtempa
    \if@nobreak \ifvmode\nobreak\fi\fi\fi\@esphack}
     \gdef\@eqnlabel{#1}}
\def\@eqnlabel{}
\def\@vacuum{}
\def\draftmarginnote#1{\marginpar{\raggedright\scriptsize\tt#1}}
\def\draft{\oddsidemargin -.5truein
        \def\@oddfoot{\sl preliminary draft \hfil
        \rm\thepage\hfil\sl\today\quad\militarytime}
        \let\@evenfoot\@oddfoot \overfullrule 3pt
        \let\label=\draftlabel
        \let\marginnote=\draftmarginnote

\def\@eqnnum{(\theequation)\rlap{\kern\marginparsep\tt\@eqnlabel}%
\global\let\@eqnlabel\@vacuum}  }
\def\preprint{\twocolumn\sloppy\flushbottom\parindent 1em
        \leftmargini 2em\leftmarginv .5em\leftmarginvi .5em
        \oddsidemargin -.5in    \evensidemargin -.5in
        \columnsep 15mm \footheight 0pt
        \textwidth 250mmin      \topmargin  -.4in
        \headheight 12pt \topskip .4in
        \textheight 175mm
        \footskip 0pt

\def\@oddhead{\thepage\hfil\addtocounter{page}{1}\thepage}
        \let\@evenhead\@oddhead \def\@oddfoot{} \def\@evenfoot{}  }
\def\titlepage{\@restonecolfalse\if@twocolumn\@restonecoltrue\onecolumn
     \else \newpage \fi \thispagestyle{empty}\c@page\z@
        \def\thefootnote{\fnsymbol{footnote}} }
\def\endtitlepage{\if@restonecol\twocolumn \else  \fi
        \def\thefootnote{\arabic{footnote}}
        \setcounter{footnote}{0}}  %\c@footnote\z@ }
\catcode`@=12
\relax
\def\ps@headings{\def\@oddfoot{}\def\@evenfoot{}
\def\@oddhead{\hbox{}\hfill
        \makebox[.5\textwidth]{\raggedright\ignorespaces --\thepage{}--
        \hfill }}
\def\@evenhead{\@oddhead}
\def\subsectionmark##1{\markboth{##1}{}} }

\ps@headings

\relax

\def\firstpage#1#2#3#4#5#6{
\begin{document}
%%%%%%%%%%%%%%%%%%%%%%%%%%%%%%%%%%%%%%%%
%\draft
%%%%%%%%%%%%%%%%%%%%%%%%%%%%%%%%%%%%%%%%

%%%%%%%%%% end of standard macros  %%%%%%%%%%%%%%%

\begin{titlepage}
\nopagebreak
\title{\begin{flushright}
        \vspace*{-1.2in}
       {\normalsize RM3-TH/05-1} \\[-6mm]
       {\normalsize Imperial/TP/050101} \\[-6mm]
%         {\normalsize hep-th/yymmddd }
\end{flushright}
\vfill {#3}}
\vskip 12pt
\author{\large #4 \\[1.0cm] #5}
\maketitle
\vskip -9mm
\nopagebreak
\vskip 48pt
\begin{abstract} {\noindent #6}
\end{abstract}
\vskip 48pt
\begin{center}
{\sl Based on the lecture presented by C.M.~Hull at the First
Solvay Workshop on Higher-Spin Gauge Theories, held in Brussels on
May 12-14, \ 2004 }
\end{center}
\vfill
\begin{flushleft}
\rule{16cm}{0.2mm}\\
\vskip 18 pt
\end{flushleft}
\thispagestyle{empty}
\end{titlepage}}

\date{}
\firstpage{3118}{IC/95/34} {\Large\bf Higher-Spin Gauge Fields and
Duality} {D.~Francia${}^a$ \ and \
C.M.~Hull${}^b$}
%\\[-4mm]
{ ${}^a$ \small\sl Dipartimento di Fisica, Universit\`a di Roma Tre, \\ [-3mm]
 \small \sl INFN, Sezione di Roma III,\\ [-3mm]
\small\sl via della Vasca Navale 84, I-00146 Roma, Italy \\
 ${}^b$ \small\sl Theoretical Physics Group, Blackett Laboratory,\\
[-3mm]
 \small \sl Imperial College of Science and Technology\\ [-3mm]
\small\sl London SW7 2BZ, U.K.} {
We review the construction of free gauge theories for gauge fields in arbitrary representations of the
Lorentz group in $D$ dimensions. We describe
the multi-form
calculus which gives the natural geometric framework for these theories.
We also discuss duality transformations that give different field theory representations of the same
physical degrees of freedom, and discuss the example of gravity in $D$ dimensions and its dual realisations in detail.
} \break

%%%%%%%%%%%%%%%%%%%%%%%%%%%%%%%%%%%%%%%%%%%%%%%%%%%%%%%%%%%%%%%%%%%%%%%%%%%%%
\section{Introduction}

Tensor fields in exotic higher-spin representations of the Lorentz group arise
as massive modes in string theory, and limits in which such fields might become massless
are of particular interest. In such cases, these would have to become higher-spin gauge fields
with appropriate gauge invariance.
Such exotic gauge fields can also arise  as dual representations of more familiar gauge theories
\cite{Hull:2000zn}, \cite{Hull:2001iu}.
The purpose here  is to review the   formulation of such exotic gauge theories that was developed in collaboration with Paul de Medeiros in
\cite{deMedeiros:2002ge}, \cite{deMedeiros:2003dc}.

Free massless particles in   $D$-dimensional  Minkowski space are classified by
representations of the little group $SO(D-2)$. A bosonic particle is associated with a tensor field
$A_{ij...k}$ in some irreducible tensor representation of $SO(D-2)$ and in physical
gauge (i.e. in  light-cone gauge) the particle is described by a field $A_{ij...k}$ depending on all $D$ coordinates of
Minkowski space and satisfying a free wave equation
\begin{equation}
\Box A =0 \ .
\end{equation}
For $D=4$, the bosonic representations of the little group $SO(2)$ are classified by an integer, the spin
 $s$, while for $D>4$ the representation theory is more involved, although it is common to  still refer to generic
 tensors  as being of \lq higher spin'.

The main topic to be considered here is the construction of the Lorentz-covariant gauge  theory
corresponding to these free physical-gauge theories.
The first step is finding the appropriate covariant tensor gauge field.
For example, an $n$'th rank antisymmetric tensor physical-gauge field $A_{i_1...i_n}= A_{[i_1...i_n]}$
(where $i,j=1,...,D-2$) arises from a covariant $n$'th rank antisymmetric tensor  gauge field $A_{\m_1...\m_n}= A_{[\m_1...\m_n]}$ (where $\m,\n=0,1,...,D-1$) with
gauge symmetry $\d A=d\l$, while a graviton
represented by a traceless  symmetric tensor
$h_{ij}=h_{ji}$ with $h_i{}^i =0$ arises from a covariant tensor gauge field
$h_{\m\n}$ which is symmetric but not traceless, with the usual gauge transformations corresponding to linearised diffeomorphisms.
The general rule is to replace an irreducible  tensor representation of $SO(D-2)$, given by some
tensor field
$A_{ij...k}$ with suitable trace-free constraints, by the corresponding  tensor field
$A_{\m\n...\rho}$ with the same symmetry properties as $A_{ij...k}$, but with no constraints on the traces, so that it can be viewed as a
tensor representation of
$GL(D,\R)$.
There are some subtleties in this step which we shall return to shortly.
The covariant gauge field must transform under gauge symmetries that are sufficient
to remove all negative-norm states and to allow the recovery of the physical-gauge theory on gauge fixing.

The next step is the construction of a gauge-invariant field equation and action.
For antisymmetric tensors or gravitons, this is straightforward, but for generic higher spin
representations the situation is more complicated.
One of the simplest cases is that of totally symmetric tensor gauge fields $A_{\m_1...\m_n}= A_{(\m_1...\m_n)}$.
For these, covariant field equations were found by Fronsdal in \cite{Fronsdal:1978rb} and reformulated in a geometric language by de Wit
and Freedman in \cite{deWit:1979pe},
but these suffered from the drawback that the gauge fields were constrained, corresponding to a partial fixing of the gauge invariance.  This was generalised to arbitrary representations  by Siegel and Zwiebach \cite{keysieg}, and the duality properties analysed.
Covariant field equations and actions
 have very
recently been constructed for totally symmetric tensor gauge fields by Francia and Sagnotti
\cite{Francia:2002aa}, \cite{Francia:2002pt} (for a review see the contribution
to these proceedings \cite{Bouatta:2004kk}). These have an elegant geometrical structure,
being constructed in terms of covariant field strengths, but have the surprising feature of being non-local in general.
Nonetheless,  on partially fixing the gauge invariance the non-locality is eliminated
and the field equations of \cite{Fronsdal:1978rb}, \cite{deWit:1979pe} are recovered.
It appears that this non-locality is inescapable in the covariant formulation
of higher-spin gauge theories, and it would be interesting to understand whether  this has any physical consequences.

Recently, this has been generalised to general  higher spin  gauge fields in any tensor representation
\cite{deMedeiros:2002ge}, \cite{deMedeiros:2003dc},  \cite{Bekaert:2002dt}, \cite{Bekaert:2003az}.
The formulation  of  \cite{deMedeiros:2002ge}, \cite{deMedeiros:2003dc}
uses an elegant mathematical structure, the multiform calculus, developed in \cite{deMedeiros:2002ge}, \cite{deMedeiros:2003dc}
and in   \cite{Dubois-Violette:1999rd},
\cite{Dubois-Violette:2000ee}, \cite{Dubois-Violette:2001jk}. It
 is the  approach of  \cite{deMedeiros:2002ge}, \cite{deMedeiros:2003dc}     which will be reviewed here.
The theory is formulated in terms of covariant field strengths or curvatures,
and is non-local but reduces to a local theory on gauge-fixing.

In general, it turns out that a given particle theory corresponding
to a particular  irreducible tensor representation of $SO(D-2)$ can arise from a number of different covariant field theories, and these
covariant field theories are said to give dual realisations of the same theory  \cite{Hull:2000zn}, \cite{Hull:2001iu}.
For example, consider an $n$-form representation of $SO(D-2)$ with field
$A_{i_1...i_n}$. This is equivalent to
the $\tilde n$-form representation, where $\tilde n=D-2-n$ and so the theory could
instead be represented in terms of an  $\tilde n$-form
field $\ti A_{i_1...i_{\ti n} } \equiv \frac {1}{n!} \e _{i_1...i_{\ti n} j_1...j_n} A^{j_1...j_n}$.
One can then construct a covariant gauge theory based on an $n$-form gauge field
$A_{\m_1...\m_n}$ or a $\ti n$-form gauge field
$\ti A_{\m_1...\m_{\ti n}}$. These are physically equivalent classically, as they both give equivalent
theories in physical gauge. The key feature here is that $n$-form and $\ti n$-form
representations are equivalent for $SO(D-2)$ but distinct for $GL(D,\R)$.
For the general case, there are a number of distinct representations of $GL(D,\R)$
that give rise to equivalent representations of $SO(D-2)$ and so lead to dual formulations of the same   physical degrees of freedom.
Such dualities  \cite{Hull:2000zn}, \cite{Hull:2001iu} can be  considered in  multi-form gauge theories
and in general interchange field equations and Bianchi identities and will also be briefly reviewed here.

\section{Young Tableaux}

Representations of $GL(D,\R)$ can be represented by Young tableaux, with each index $\m$ of a tensor $T_{\m \n .... \rho }$ corresponding to a box in the diagram; see
\cite{ham} for a full discussion.  Symmetrized indices are represented by boxes arranged in a row, so that e.g.
a 2nd rank symmetric tensor $h_{\m\n}$  is represented by \, {\tiny \begin{tabular}{|c|c|}
\hline
& \\
\hline
\end{tabular}}\ ,
while anti-symmetrized indices are represented by boxes arranged in a column, so that e.g.
a 2nd rank anti-symmetric tensor $B_{\m\n}$  is represented by
\, {\tiny \begin{tabular}{|c|}
\hline
\\
\hline
\\
\hline
\end{tabular}} \ .
A general 3rd rank tensor $E_{\m\n\rho}$ can be decomposed into a totally symmetric piece
$E_{(\m\n\rho)}$ represented by the tableau
\,
{\tiny \begin{tabular}{|c|c|c|}
\hline
& & \\
\hline
\end{tabular}}\, ,
a totally anti-symmetric piece
$E_{[\m\n\rho]}$ represented by the tableau
\, {\tiny \begin{tabular}{|c|}
\hline
\\
\hline
\\
\hline
\\
\hline
\end{tabular}} \, ,
and the remaining piece $D_{\m\n\rho}\equiv
E_{\m\n\rho}-
E_{(\m\n\rho)}
-E_{[\m\n\rho]}
$, which is said to be of  mixed symmetry,
  is represented by the
``hook'' tableau:
\ {\tiny \begin{tabular}{|c|c|}
\hline
& \\
\hline
\\
\cline{1-1}
\end{tabular}}\ .
This   satisfies $D_{[\m\n\rho]}=0$
  and $D_{(\m\n\rho)}=0$
and  is an irreducible representation of $GL(D,\R)$.
As another example, a fourth-rank tensor $R_{\m\n\rho\s}$ with the symmetries of the Riemann tensor
corresponds to the diagram
\ {\tiny \begin{tabular}{|c|c|}
\hline
& \\
\hline
& \\
\hline
\end{tabular}} .

The same diagrams can be used also  to classify representations of $SO(D)$, but with the difference that now
all traces must be removed to obtain an irreducible representation.
For example, the diagram
 \ {\tiny \begin{tabular}{|c|c|}
\hline
& \\
\hline
\end{tabular}}
now regarded as a tableau for $SO(D)$ corresponds to
2nd rank symmetric tensor $h_{\m\n}$ that  is traceless,
 $\d^{\m\n}h_{\m\n}=0$.
 The hook tableau
\ {\tiny \begin{tabular}{|c|c|}
\hline
& \\
\hline
\\
\cline{1-1}
\end{tabular}}
now corresponds to a tensor $D_{\m\n\rho}$ that is traceless,   $\d^{\m\rho}D_{\m\n\rho}=0$.
Similarly,  the diagram
\ {\tiny \begin{tabular}{|c|c|}
\hline
& \\
\hline
& \\
\hline
\end{tabular}}  now corresponds to a tensor with the algebraic properties of the Weyl  tensor.

Then given a field    in physical gauge   in a representation of $SO(D-2)$ corresponding to some Young tableau, the corresponding covariant field
in the construction outlined above is
 in the representation of $GL(D,\R) $ corresponding to the {\it same} Young tableau, now regarded as a tableau for $GL(D,\R) $.
 For example, a graviton is represented in physical gauge by a transverse traceless tensor $h_{ij}$ (with
 $\d ^{ij}h_{ij}=0$) of
  $SO(D-2)$ corresponding to the Young tableau \, {\tiny \begin{tabular}{|c|c|}
\hline
& \\
\hline
\end{tabular}}\ ,
so the covariant formulation is the
 $GL(D,\R) $ representation with tableau
  \, {\tiny \begin{tabular}{|c|c|}
\hline
& \\
\hline
\end{tabular}}\ , which is a symmetric tensor $h_{\m\n}$ with no constraints on its trace.

 It will be convenient to label tableaux by the lengths of their columns, so that
 a tableau  with columns of length $n_1,n_2,...,n_p$ will be said to be of type $[n_1,n_2,...,n_p]$.
  It is conventional to
 arrange these in decreasing order, $n_1\ge n_2\ge ... \ge n_p$.

\section{Duality}

Free gauge theories typically have a number of dual formulations. For example, electromagnetism
in flat $D$ dimensional space is formulated in terms of a 2-form field strength $F = \12 F_{\m\n} \ dx^{\m}
\wedge \ dx^{\n}$
satisfying $dF= 0$ and $  \ d \ \ast  {F} = 0$, where $\ast F$ denotes the
Hodge dual $D-2$ form with components
\be
 \ast
F_{\m_1...\m_{D-2} }\equiv   \12 F^{  \n\rho} \e_{ \n\rho \m_1...\m_{D-2}} \ .
\ee
The equation $dF= 0$ can be interpreted as a Bianchi identity and solved in terms of a
1-form potential $A$ as $F=dA$, with $ \ d \ \ast  {F} = 0$ regarded as a field equation for $A$.
Alternatively, one can view $  \ d \ \ast  {F} = 0$ as the Bianchi identity $d\tilde{F} =0$ for $\tilde{F} \equiv \ast
F$, and this
  implies that $\tilde F$ can be written   in terms of a $D-3$ form potential $\tilde{A}$ with
$\tilde{F} = d\tilde{A}$. Then $dF= 0$ becomes $ d\ast \tilde F=0$ which can be
regarded as a field equation for $\tilde A$.
The theory can be formulated either  in terms of the one-form $A$ or in terms of  the $D-3$ form potential $\tilde{A}$,
giving two dual formulations.

This can be understood from the point of view of the little group $SO(D-2)$.
In physical gauge or  light-cone gauge, the degrees of freedom are represented by a transverse vector field
$A_i$ in the $D-2$ dimensional vector representation of  $SO(D-2)$, with $i=1 \dots D-2$.
This is equivalent to the $(D-3)$-form representation of $SO(D-2)$,
so the theory can equivalently be formulated in physical gauge in terms of a
$(D-3)$-form
\be
\label{dul}
\tilde{A}_{j_{1}\cdots j_{n}}=\e_{j_{1}\cdots j_{n}i}A^{i} \ .
\ee
where $n=D-3$.
These representations of $SO(D-2)$  can be associated with Young tableaux. The vector representation of $SO(D-2)$
 is described
by a single-box Young tableau,
\ {\tiny \begin{tabular}{|c|}\hline
\\ \hline
\end{tabular}} \ ,
while the $(D-3)$-form
is associated with a tableau  that has
one column of $D-3$ boxes. For example in $D=5$, this
 is a
one-column, two-box tableau,
\ {\tiny \begin{tabular}{|c|}
\hline
\\
\hline
\\
\hline
\end{tabular}} \ .

In physical gauge, changing from a 1-form field $A_i$ to a
$D-3 $ form field $\tilde{A}_{j_{1}\cdots j_{n}}$ is the local field redefinition (\ref{dul})
and so is a trivial rewriting of the theory. However, these lead to  two different
formulations of the covariant theory: the same physical degrees of freedom can be obtained either from a
covariant 1-form gauge field $A_\m$ transforming as a vector under $SO(D-1,1)$, or from a
$D-3 $ form gauge field $\tilde{A}_{\m _{1}\cdots \m_{n}}$. The one-form field has
a gauge symmetry $\d A=d\l$ while the $D-3$ form field has a gauge  symmetry
$\d \tilde A=d \tilde \l$ and these can be used  to   eliminate  the unphysical
degrees of freedom and go to physical gauge.
Thus two formulations that are equivalent in  physical gauge correspond to two covariant formulations
that are distinct covariant realisations of the theory.

This is the key to understanding the generalisations to other gauge fields
in other representations of the Lorentz group. A scalar field is a singlet of the
little group, and this is equivalent to the $D-2$ form representation of $SO(D-2)$,
represented by a tableau with one  column consisting of $D-2$ boxes.
The scalar field $\phi$ then has a dual covariant  formulation
as  a $D-2$ form gauge field $\tilde{\phi}_{\m _{1}\cdots \m_{D-2}}$.

 For spin $2$, the graviton in $D$ dimensions is a field $h_{\m\n}$ which is a
 symmetric second rank (with trace)
 represented by the Young tableau \ {\tiny \begin{tabular}{|c|c|}
\hline
& \\
\hline
\end{tabular}}
for $GL(D,\R)$.
The reduction to physical gauge  gives a transverse, symmetric, traceless tensor of $SO(D-2)$
$h_{ij}$, corresponding to the $SO(D-2)$ tableau
\ {\tiny \begin{tabular}{|c|c|}
\hline
& \\
\hline
\end{tabular}}  .
(Recall that for $GL(D,\R)$, each box in the tableau represents an index in the $D$-dimensional representation
and   has a trace in general, while for
 $SO(D-2)$ each box in the tableau represents an index in the $(D-2)$-dimensional representation
and traces are removed using the $SO(D-2)$ metric, so that the symmetric tensor
$h_{ij}$ satisfies
$\d ^{ij}h_{ij}=0$.)
The physical gauge graviton $h_{ij}$
can be dualized on one or both of its indices
 giving respectively
\ba
D_{i_{1}\cdots i_{n}k}                 & = &\e_{i_{1}\cdots i_{n}l} \ h^{l}_{\  k} \ , \\
  \label{dull}
C_{i_{1}\cdots i_{n}j_{1}\cdots j_{n}} & = &\e_{i_{1}\cdots
i_{n}l} \ \e_{j_{1}\cdots j_{n}k} \ h^{lk} \ .
\ea
These give equivalent representations of the little group
$SO(D-2)$ , with   appropriate trace conditions.
  The tracelessness condition
$\d ^{ij}h_{ij}=0$ implies $D_{[i_{1}\cdots i_{n}k]} =0$, while the symmetry
$h_{[ij]}=0$ implies the tracelessness
$\d ^{ i_{n}k}D_{i_{1}\cdots i_{n}k}=0$.
Then $D$ is   represented by the $[n,1]$  hook diagram with one column of length $n=D-3$ and one of length one, so that in dimension $D=5$, $D_{ijk}$ corresponds to the   ``hook'' tableau for $SO(D-2)$:
\ {\tiny \begin{tabular}{|c|c|}
\hline
& \\
\hline
\\
\cline{1-1}
\end{tabular}} .
The field
   $C_{i_{1}\cdots i_{n}j_{1}\cdots j_{n}}$ corresponds to the
  tableau for $GL(D-2,\R)$ of type $[n,n]$ with two columns each of $n=D-3$ boxes, so that for $D=5$ $C_{ijkl}$ corresponds to  the
  ``window", the two-times-two tableau:
\ {\tiny \begin{tabular}{|c|c|}
\hline
& \\
\hline
& \\
\hline
\end{tabular}} .
  However, it turns out that  $C_{i_{1}\cdots i_{n}j_{1}\cdots j_{n}}$ is not in the $[n,n]$ representation for $SO(D-2)$.
  In general, the
  $[m,m]$ representation of $GL(D-2,\R)$ would decompose into
  the representations $[m,m] \oplus  [m-1,m-1] \oplus [m-2,m-2] \oplus .... $ of $SO(D-2)$,
  corresponding to multiple traces. For $m=n=D-3$, it turns out that all the trace-free parts vanish identically, so that only the $[1,1]$ and singlet representations of  $SO(D-2)$ survive resulting from $n-1$ and $n$ traces respectively, so that
   $$C_{i_{1}\cdots i_{n}}{}^{j_{1}\cdots j_{n}}= \d _{[i_{1}}{}^{[j_{1}}  \cdots
    \d _{i_{n-1}}{}^{j_{n-1}}   C _{i_{n}]}{}^{j_{n}]} +\d _{[i_{1}}{}^{[j_{1}}  \cdots
    \d _{i_{n-1}}{}^{j_{n-1}}   \d _{i_{n}]}{}^{j_{n}]}C$$
    for some    $C_{ij},C$ with traceless $C_{ij}$.
    The definition  (\ref{dull})
    and the tracelessness of $h_{ij}$ then imply that taking $n$ traces of
    $C_{i_{1}\cdots i_{n}j_{1}\cdots j_{n}}$ gives zero, so that $C=0$ and  $C_{ij}$ is traceless and in the representation $[1,1]$, and in fact $C_{ij}$ is proportional to $h_{ij}$.

For
arbitrary spin in dimension $D$ the general form for a gauge field
in light-cone gauge will be
$$
D_{[i_{1}\cdots i_{n_{1}}][j_{1}\cdots j_{n_{2}}]\cdots } \ ,
$$
corresponding to an arbitrary representation of the little group
$SO(D-2)$, described by a Young tableau with an arbitrary number
of columns of lenghts $ n_{1},n_{2} \cdots $:
{\begin{center}
 {\tiny \begin{tabular}{|c|c|c|c|c|}
\multicolumn{5}{c}{$n_1$ \ $n_2$ \ $n_3$ \, \   \dots \ \, \, $n_p$} \\
 \cline{1-5}
  & & & &  \dots  \\
\cline{1-5}
  & & & $\ddots$ \\
\cline{1-4}
  & & \vdots \\
\cline{1-3}
  &  \vdots \\
\cline{1-2}
\vdots \\
\cline{1-1}
\end{tabular}}
\end{center}}
Dual descriptions of
such fields can be obtained by dualising any column, i.e. by
replacing one of length $m$ with one of length $D-2-m$  (and re-ordering the sequence of columns, if necessary),
or by simultaneously dualising a number of columns
\cite{Hull:2001iu}. Then any of the equivalent physical gauge fields can be covariantised to a
gauge field associated with the same tableau, but now viewed as defining a representation of $GL(D,\R)$.
The set of Young tableaux for these dual representations of the same theory
define distinct representations of $GL(D,\R)$, but all reduce to equivalent representations of the little group $SO(D-2)$.

In fact, there are yet further dual representations.
For $SO(D-2)$, a column of length $D-2$ is a singlet, and
given any tableau for $SO(D-2)$, one can obtain yet more dual formulations
by adding any number of columns of length $D-2$, then reinterpreting as a
tableau for $GL(D,\R)$ \cite{keysieg}. Thus for a vector field in $D=5$, there are
dual representations with gauge fields in the representations of $GL(D,\R)$ corresponding to the following tableaux:

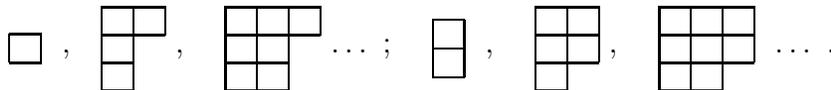
\begin{center}
%1
\, {\tiny \begin{tabular}{|c|}\hline
\\ \hline
\end{tabular}} \ ,\,
%
%3-1
\, {\tiny \begin{tabular}{|c|c|}
\cline{1-2}
& \\
\cline{1-2}
\\
\cline{1-1}
\\
\cline{1-1}
\end{tabular}}\ , \,
%
%3-3-1
\, {\tiny \begin{tabular}{|c|c|c|}
\cline{1-3}
& & \\
\cline{1-3}
& \\
\cline{1-2}
& \\
\cline{1-2}
\end{tabular}}\ \dots \ ; \,
%2
\, {\tiny \begin{tabular}{|c|}
\hline
\\
\hline
\\
\hline
\end{tabular}} \ , \,
%
%3-2
\, {\tiny \begin{tabular}{|c|c|}
\cline{1-2}
& \\
\cline{1-2}
& \\
\cline{1-2}
\\
\cline{1-1}
\end{tabular}}\ , \,
%
%3-3-2
\, {\tiny \begin{tabular}{|c|c|c|}
\cline{1-3}
& & \\
\cline{1-3}
& & \\
\cline{1-3}
& \\
\cline{1-2}
\end{tabular}} \ \dots \ .
\end{center}

\section{Bi-forms}

Before turning to general gauge fields in general representations, we consider
the simplest new case, that  of gauge fields in representations corresponding to Young
tableaux with two columns. It is  useful  to consider first
bi-forms, which are reducible representations in general, arising from the tensor
product of two forms, and then at a later stage project onto the
irreducible representation corresponding to a Young tableau with two columns.
In this section we review the calculus for bi-forms of \cite{deMedeiros:2002ge}
and generalise to multi-forms and general tableaux in section $7$.

A \emph{bi-form}  of type $(p,q)$ is an element $T$ of $X^{p,q}$, where
  $X^{p,q}\equiv \L^{p}\otimes\L^{q}$ is the
 $GL(D,\mathbb{R})$ -~reducible
tensor product of
the space $ \L^{p}$ of $p$-forms with the space $ \L^{q}$ of
$q$-forms on $\mathbb{R}^{D}$.
In components: \be
T=\fr{1}{p!\, q!}T_{\m_{1}\cdots\m_{p}\n_{1}\cdots\n_{q}} dx^{\m_{1}} \ww
\cdots \ww dx^{\m_{p}}\otimes dx^{\n_{1}}\ww \cdots \ww
dx^{\n_{q}} \ . \ee
and is specified by a tensor $T_{\m_{1}\cdots\m_{p}\n_{1}\cdots\n_{q}} $
which is
antisymmetric on each of the two sets of $p$ and $q$ indices
$T_{\m_{1}\cdots\m_{p}\n_{1}\cdots\n_{q}} =T_{[\m_{1}\cdots\m_{p}][\n_{1}\cdots\n_{q}]} $, and no
other symmetries are assumed.
One can define a number of operations on bi-forms: here we only
describe the ones needed for the forthcoming discussion, referring
to \cite{deMedeiros:2002ge} for a more complete development.

 Two exterior derivatives, acting on the two sets of indices, are
defined as

\hspace{3cm}$d:X^{p,q}\ra X^{p+1,q}\ ,$ \hspace{4cm} \emph{left derivative}

\hspace{3cm}$\tilde{d}:X^{p,q}\ra X^{p,q+1} \ ,$ \hspace{4cm} \emph{right derivative}
whose action on the elements of $X^{p,q}$ is
\be
\begin{split}
d\, T & =  \fr{1}{p!\,q!}\pr_{[\m}T_{\m_{1}\cdots\m_{p}]\n_{1}\cdots\n_{q}} dx^{\m} \ww dx^{\m_{1}}\ww \cdots \ww dx^{\m_{p}}
\otimes dx^{\n_{1}}\ww \cdots \ww dx^{\n_{q}} \ ,\\
\tilde{d}\, T & =  \fr{1}{p!\,q!}\pr_{[\n}T_{|\m_{1}\cdots\m_{p}|\n_{1}\cdots\n_{q}]}dx^{\m_{1}}\ww \cdots \ww dx^{\m_{p}}
\otimes dx^{\n} \ww  dx^{\n_{1}}\ww \cdots \ww dx^{\n_{q}} \ ,
\end{split}
\ee
where the two sets of antisymmetric indices are separated by
vertical bars. One can verify that
\be
d^{2}=0=\tilde{d}^{2},
\hspace{3cm} d\,\tilde{d}=\tilde{d}\,d \ .
\ee
With the help of these
two exterior derivatives, one can also define the \emph{total
derivative}
\ba
{\cal{D}} \equiv  d + \tilde{d} \ , \hspace{2cm}
\mbox{such that} \hspace{2cm} {\cal{D}}^{3} = 0 \ ,
\ea
where the nilpotency of $\cal{D}$ is a straightforward consequence of the nilpotency of
$d$ and $\tilde{d}$. Such nilpotent differential operators were considered by
  \cite{Dubois-Violette:1999rd},\cite{Dubois-Violette:2000ee},\cite{Dubois-Violette:2001jk}.
In a
similar fashion, restricting to reducible representations of
$SO(D-1,1)$, one can introduce two distinct Hodge-duals:

\hspace{3cm}$\ast :X^{p,q}\ra X^{D-p,q} \ ,$ \hspace{4cm} \emph{left dual}

\hspace{3cm}$\tilde{\ast}:X^{p,q}\ra X^{p,D-q} \ ,$ \hspace{4cm} \emph{right dual}

defined as
\be \label{hodge}
\begin{split}
\ast \,T & =   \fr{1}{p!\,(D-p)!\,q!}{T_{\m_{1}\cdots\m_{p}\n_{1}\cdots\n_{q}}
\e^{\m_{1}\cdots\m_{p}}_{\ \ \ \ \ \ \m_{p+1}\cdots\m_{D}}
dx^{\m_{p+1}}\ww \cdots \ww dx^{\m_{D}}
\otimes dx^{\n_{1}}\ww \cdots \ww dx^{\n_{q}}} \ , \\
\tilde{\ast} \,T & =
\fr{1}{p!\,q!\,(D-q)!}{T_{\m_{1}\cdots\m_{p}\n_{1}\cdots\n_{q}}
\e^{\n_{1}\cdots\n_{q}}_{\ \ \ \ \ \ \n_{q+1}\cdots\n_{D}}
dx^{\m_{1}}\ww \cdots \ww dx^{\m_{p}} \otimes dx^{\n_{q+1}}\ww
\cdots \ww dx^{\n_{D}}} \ .
\end{split}
\ee
 These definitions imply that
\be \label{gravidual}
\ast^{2}=(-1)^{1+p(D-p)},\hspace{1.5cm}
\tilde{\ast}^{2} = (-1)^{1+q(D-q)} \ , \hspace{1.5cm} \ast
\tilde{\ast}=\tilde{\ast} \ast \ ,
\ee
as can be verified recalling
the contraction identity for the Ricci-tensor in $D$ dimensions:
\be \label{ricci}
\e^{\a_1 \dots \a_k \a_{k+1} \dots \a_D} \e_{\a_1 \dots \a_k
\b_{k+1} \dots \b_D} \ = \ -(D-k)! \cdot
(\d^{[\a_{k+1}}_{\b_{k+1}} \dots \d^{\a_D]}_{\b_D}) \ ,
\ee
where we are using the ``mostly plus'' flat background metric.

There are
  three operations on bi-forms that   enter the Bianchi
identities and the equations of motion, and into the projections onto irreducible representations: a trace, a dual trace, and
a transposition.

A trace operator acts on a pair of indices belonging to different sets,
so that
$$
\t : X^{p,q} \ra X^{p-1,q-1} \ ,
$$
and is defined   by
\be
\t \,T \equiv
\fr{1}{(p-1)!\,(q-1)!}\h^{\m_1\n_1}T_{\m_{1}\cdots\m_{p}\n_{1}\cdots\n_{q}}dx^{\m_{2}}\ww
\cdots \ww dx^{\m_{p}} \otimes dx^{\n_{2}}\ww \cdots \ww
dx^{\n_{q}} \ .
\ee
Combining the $\t$ operator with the Hodge
duals, one can also define two distinct dual traces:
\be
\begin{split}
\s         & \equiv (-1)^{1+D(p+1)} \ast \t \ast :X^{p,q}\ra X^{p+1,q-1} \ , \\
\tilde{\s} & \equiv (-1)^{1+D(q+1)} \tilde{\ast}  \t \tilde{\ast} :X^{p,q}\ra X^{p-1,q+1} \ ,
\end{split}
\ee
that antisymmetrize one index in a set with respect to the
whole other set:
\be \label{sigma}
\begin{split}
\s \,T & =  \fr{(-1)^{p+1}}{p!\,(q-1)!}T_{[\m_{1}\cdots\m_{p}\n_{1}]\cdots\n_{q}}dx^{\m_{1}}\ww \cdots \ww dx^{\m_{p}}
\otimes dx^{\n_{1}}\ww \cdots \ww dx^{\n_{q}} \ , \\
\tilde{\s} \,T & =  \fr{(-1)^{q+1}}{(p-1)!\,q!}T_{\m_{1}\cdots[\m_{p}\n_{1}\cdots\n_{q}]}dx^{\m_{1}}\ww \cdots \ww dx^{\m_{p}}
\otimes dx^{\n_{1}}\ww \cdots \ww dx^{\n_{q}} \ .
\end{split}
\ee
Again, the proof of (\ref{sigma}) relays on the identity (\ref{ricci}).
For example, for a $(2,3)$ form in $D=5$ one has (omitting combinatorial factors):
\ba
\ast \,T          & \sim & T_{p_1p_2,q_1q_2q_3} \e^{p_1p_2}_{\ \ \ \ \ p_3p_4p_5} \ , \nonumber \\
\t \ast \,T       & \sim & T_{p_1p_2,q_1q_2q_3} \e^{p_1p_2q_1}_{\ \ \ \ \ \ p_4p_5} \ ,\nonumber \\
\ast  \t \ast \,T & \sim & T_{p_1p_2,q_1q_2q_3} \e^{p_1p_2q_1}_{\ \ \ \ \ \ p_4p_5} \e^{p_4p_5}_{\ \ \ \ \a_1\a_1\a_3}
                 \sim  T_{p_1p_2,q_1q_2q_3} \d^{[p_1p_2q_1]}_{\a_1\a_2\a_3} \ .
\ea
 Finally, the transposition operator simply interchanges the
two sets of indices:
$$
t : X^{p,q} \ra X^{q,p} \ ,
$$
so that
$$
(t\,T)_{\n_{1}\cdots\n_{q}\m_{1}\cdots\m_{p}}=T_{\m_{1}\cdots\m_{p}\n_{1}\cdots\n_{q}}$$
and
\ba
t\,T \equiv
\fr{1}{p!q!}T_{\n_{1}\cdots\n_{q}\m_{1}\cdots\m_{p}}dx^{\n_{1}}\ww
\cdots \ww dx^{\n_{q}} \otimes dx^{\m_{1}}\ww \cdots \ww
dx^{\m_{p}} \ .
\ea

The bi-forms are a reducible representation of $GL(D,\R)$.
It is useful to introduce the Young symmetrizer ${\cal{Y}}_{[p,q]}$
which projects a bi-form $T$ of type $(p,q)$ onto the part $\hat T= {\cal{Y}}_{[p,q]}\,T$
lying in the irreducible representation corresponding to a tableau of type $[p,q]$,
with two  columns of length $p$ and $q$, respectively (we use round brackets for reducible
$(p,q)$ bi-forms and square ones for irreducible representations).
The projected part $\hat T$ satisfies the additional constraints (for $p\ge q$):
\be \label{const}
\begin{split}
\s \,\hat T & = 0 \ ,\\
t \,\hat T  & =  \hat T , \hspace{1cm}\mbox{if} \hspace{.5cm}  p = q \ .
\end{split}
\ee

\section{  $D$-Dimensional Linearised Gravity}

It is straightforward to
formulate gauge field theories of bi-forms;  a    gauge field
$A$ in the space $X^{p,q}$  can be thought of as a  linear combination of terms arising from the
tensor product of a $p$-form gauge field and a $q$-form gauge field.
It transforms under the gauge transformation
\ba
\label{gaugetr} \d \, A = d \, \a^{\, p-1,q} + \tilde{d} \,
\a^{\,p,q-1},
\ea
with gauge parameters that are themselves
bi-forms in $X^{\, p-1,q}\ ,\ X^{\,p,q-1}$.
 Clearly,
\ba
F = d \,\tilde{d}\, A
\ea
is a gauge invariant field-strength for $A$.
This is a convenient starting point for describing gauge fields in irreducible representations.
We now show how to project the bi-form gauge theory using Young projections to obtain irreducible
gauge theories, starting
with one of the simplest examples, that of linearised gravity in $D$
dimensions.

The graviton field is a rank-two tensor in an irreducible representation  of
$GL(D,\mathbb{R})$  described by a Young
tableau
 of type $[1,1]$, i.e. a
two-column, one-row Young
tableau,
\, {\tiny \begin{tabular}{|c|c|}
\hline
& \\
\hline
\end{tabular}} \, .
 The starting
point in our case is thus a bi-form $h \in X^{1,1}$,
corresponding to a 2nd rank tensor $h_{\m\n}$, and
 we would
like to project on the $GL(D,\mathbb{R})-$irreducible tensor of type
$[1,1]$: $\widehat h= {\cal{Y}}_{[1,1]}\,h$
using the Young projector ${\cal{Y}}_{[1,1]}$.
Then the constraints (\ref{const}) become
\be
\begin{split}
\s \,\widehat h & = 0 \ , \\
t \,\widehat h  & = \widehat h \ .
\end{split}
\ee
In this case, the two conditions  are equivalent and simply imply that $\widehat h $ is symmetric,
 $\widehat h _{\m\n}=h_{(\m\n)}$.
 The gauge transformation for the graviton is
the Young-projection of
(\ref{gaugetr})\footnote{Note that acting on a tensor in an irreducible
representation  with $d$ or $\tilde{d}$ gives  a
\emph{reducible} form in general, so that a Young projection is necessary in
order to obtain irreducible tensors.},
which gives $\d \,\widehat h _{\m\n} = \partial _{(\m} \l_{\n)}$ where
$\l _\m = \a_\m^{1,0}+   \a_\m^{0,1}$.
 The invariant field strength
is given by \footnote{The operator $d\,\tilde{d}$,
unlike $d$ and $\tilde{d}$ separately, sends irreps to
irreps, so that $\tilde{d} \,d \, \widehat{h} ={\cal{Y}}_{[1,1]} \,\tilde{d} \,d \, \widehat{h} $
and  the Young projection is automatically
implemented.}
\ba
R = \tilde{d} \,d \, \widehat{h} \ .
\ea
This is the $[2,2]$ Young tableau
\, {\tiny \begin{tabular}{|c|c|}
\hline
& \\
\hline
& \\
\hline
\end{tabular}} \,
 describing the linearized
Riemann tensor. The nilpotency of the exterior derivatives and the
irreducibility imply that the Bianchi identities
\ba
d\,R = 0 \ , \hspace{3cm} \tilde{d}\,R = 0 \ ,
\ea
\ba
\s\,R = 0 \ ,
\ea
are satisfied,
while acting with the $\t$ operator gives the Einstein equation in
$D \geq 4$
\footnote{In $D=3$ the field equation $R_{\m \n}=0$ implies
$R_{\m \n \9 \s}=0$ which only has trivial solutions; a non trivial
equation is instead $\t^{2}\,R=0$, with $\t^{2}\,R$ the Ricci-scalar \cite{Hull:2001iu}, \cite{Hull:2000zn}.
This can be generalized to (p,q)-forms, as we shall see later.} :
\label{foot}
\ba
\t\,R =   0 \ .
\ea
or in components, $R_{\m \n} = 0$.

We now return to the issue of duality.
In Section 3 we described the triality of
linearised gravity in $D$ dimensions, for which there are three different fields that can be used for describing the degrees of freedom of the graviton. The discussion can be expressed succinctly in terms of bi-forms.
In light-cone gauge the fields are tensors in irreducible representations of $SO(D-2)$, and so are trace-less.
The graviton arises from projecting a $[1,1]$ form $h$ onto a symmetric tensor $\widehat{h}_{ij}$
that is
traceless $\widehat{h}_{\ i}^{i} = 0$.
 Now one can easily
dualise the field $\widehat{h}$ in one or both indices,
by applying  (\ref{hodge}), where the $\ast$-operator is now the
 $SO(D-2)$-covariant dual. The dual light-cone   fields are
\ba
 {D} &=  \ast \,\widehat{h}  \ , \\
 {C} &=  \ast \, \tilde{\ast} \, \widehat{h} \ ,
\ea
and all have the same number of
independent components.

In the  covariant theory one dualises the field strengths rather than the gauge fields
and this is easily analysed using
the bi-form  formalism developed.
Indeed, starting from the $[2,2]$
field strength  $R$ one can define the Hodge duals
\ba
S & \equiv \ast \,R \, ,\\
G & \equiv \ast \,\tilde{\ast} \,R \, ,
\ea
which are respectively  of type $[D-2,2]$ and $[D-2,D-2]$,
associated with the tableaux:

{\begin{center}
\, {\tiny \begin{tabular}{c|c|c|}
\cline{2-3}
& & \\
\cline{2-3}
& & \\
\cline{2-3}
$D-2$ & \vdots \\
\cline{2-2}
\end{tabular}} \ \ \ \ \  and
\, {\tiny \begin{tabular}{c|c|c|c}
\cline{2-3}
& & & \\
\cline{2-3}
& & & \\
\cline{2-3}
$D-2$ & \vdots & \vdots & .\\
\cline{2-3}
\end{tabular}} \,
\end{center}}

The other possible dual, $\tilde{S} \equiv \tilde{\ast} \,R$ is
not independent, since $\tilde{S} = t\,S$ (this would not be the
case for the generalisation to a general $(p,q)$-form with $p\ne q$). \label{dual2} In components:
\ba
S_{\m_1\cdots\m_{D-2}\n_1\n_2}           & = & \fr{1}{2}
R^{\a\b} _{\ \ \n_1\n_2} \e_{\a\b\m_1\cdots\m_{D-2}} \ ,\\
G_{\m_1\cdots\m_{D-2}\n_1\cdots\n_{D-2}} & = & \fr{1}{4}
R^{\a\b\g\d}\e_{\a\b\m_1\cdots\m_{D-2}}\e_{\g\d\n_1\cdots\n_{D-2}}
\ .
\ea
From these definitions, and the
algebraic and dynamical constraints satisfied by the linearised Riemann
tensor $R$,
one can deduce a set of relations
that must be obeyed by the bi-forms $S$ and $G$.
We will give examples of
certain relations between Bianchi identities and equations of
motion, referring to \cite{deMedeiros:2002ge}, \cite{Hull:2001iu},
\cite{Hull:2000zn}
for a more complete
discussion.  The definitions $S
\equiv \ast \,R$ and $G \equiv \ast \,\tilde{\ast} \,R$, and the
relations (\ref{gravidual}) imply that $\ast \,S = (-1)^D R$, and $\tilde{\ast}\,\ast\,G = R$.
Then, using the definitions given in Section 4,
it follows that
\be
\s \,R = 0 \hspace{.3cm} \Rightarrow
\hspace{.3cm} \s \,\ast \,S = 0 \hspace{.3cm} \Rightarrow
\hspace{.3cm} \ast \,\s \,\ast \,S = 0 \hspace{.3cm}  \Rightarrow
\hspace{.3cm} \t \,S = 0 \ ;
\ee
\be
\s R = 0 \hspace{.2cm} \Rightarrow
\hspace{.2cm} \s \,\tilde{\ast}\,\ast \,G = 0 \hspace{.2cm} \Rightarrow
\hspace{.2cm} \ast \,\s \,\tilde{\ast}\,\ast \,G = 0 \hspace{.2cm}
\Rightarrow \hspace{.2cm} \t \,\tilde{\ast} \,G = 0 \hspace{.2cm}
\Rightarrow \hspace{.2cm} \tilde{\ast} \,\t \tilde{\ast} \,G = 0
\hspace{.2cm} \Rightarrow \hspace{.2cm} \tilde{\s} \,G = 0 \ .
\ee
That is to say, the Bianchi identity $\s \,R = 0$ for $R$ implies the equation
of motion $\t \,S = 0$ for $S$ and  the Bianchi identity $ \tilde{\s} \,G = 0$ for $G$. The
equation of motion $\t \,R = 0$ for  $R$ in $D>3$ implies that
\be
\t \,R = 0
\hspace{.3cm} \Rightarrow \hspace{.3cm} \ast \,\t \,\ast \,S = 0
\hspace{.3cm} \Rightarrow \hspace{.3cm} \s \,S = 0 \ ,
\ee
and
\be
\t \,R
= 0 \hspace{.3cm} \Rightarrow \hspace{.3cm} \t \,\tilde{\ast}\, \ast \,G
= 0 \hspace{.3cm} \Rightarrow \hspace{.3cm} \t^{D-3} \,G = 0 \ .
\ee
giving the Bianchi identity $\s \,S = 0 $ for $S$ and the field
equation $\t^{D-3} \,G = 0$ for $G$ \footnote{Note that
the equation  $\t^{n} \,G = 0$ only has trivial solutions for $n<D-3$, so that this is the simplest
non-trivial field equation \cite{Hull:2001iu}. }.

Other consequences for $S$ and $G$ can be deduced starting from
properties of $R$ and making use of identities involving the
various bi-form operators   (see
\cite{deMedeiros:2002ge}).
 In particular the Bianchi identities
 $d\,S=\ti d\,S=0$ and  $d\,G=\ti d\,G=0$
imply that $S$ and $G$ can   be expressed as
field-strengths of   gauge potentials $\widehat D$ and $\widehat C$ respectively, which are in  irreducible representations
 of type $[D-3,1]$ and $[D-3,D-3]$
\begin{equation}
S=d \,\ti d \,\widehat D, \qquad G=d \,\ti d \,\widehat C \ ,
\end{equation}
whose linearized equations of motion are $\t \,S = 0$ and
$\t^{D-3} \,G = 0$.
Although these relations can be derived for gravity straightforwardly, as in \cite{Hull:2001iu}, the
bi-form formalism   simplifies the discussion
and generalises to general multi-form representations
in a way that elucidates the geometric structure and allows simple derivations and calculations.

\section{General Bi-Form Gauge Theories}

The discussion of gravity  extends straightforwardly to arbitrary
$(p,q)$-forms, where without loss of generality we assume $p\ge q$.  First, one can restrict from a $(p,q)$-form $T$ to
$\widehat T= {\cal Y }_{[p,q]}\,T$
which is in $[p,q]$ irrep of $GL(D,\mathbb{R})$ satisfying  the constraints
(\ref{const}).
Then one can
define a field strength $F \equiv \tilde{d}\, d \,\widehat{T}$ of type $[p+1,q+1]$ that
is invariant under the gauge transformations
given by the projection of (\ref{gaugetr}):
\be
\d \widehat T=  {\cal{Y}}_{[p,q]} \,(d \, \a^{\, p-1,q} + \tilde{d} \,\a^{\,p,q-1}),
\ee
and satisfies the Bianchi identities
\be
d\,F=\ti d \,F=0 \
, \qquad
 \s \,F=0 \ ,
 \ee
 together with $t\,F=F$ \ if \  $p=q$.
We now turn to the generalisation of the
``Einstein equation''
$\t \,R = 0$.
The natural guess is
 \be \t \,F = 0\ , \ee
However, we have seen that
for gravity in $D=3$, the Einstein  equation $\t \,R = 0$ is too strong and only has trivial solutions,
but that the weaker condition $\t ^2\,R = 0$ (requiring that the Ricci scalar is zero)
gives a non-trivial theory.
For the dual field strength $G$, the field equation was $\t^{D-3} \,G = 0$ in $D$ dimensions.
Then it is to be expected that the
``Einstein equation''
$\t \,R = 0$ will be generalised to $[p,q]$ forms by taking
$ \t \,F = 0$ for large enough space-time dimension $D$, but for low $D$
a   number of traces of the field
strength
may be needed to give an
equation of motion
$ \t^n \,F = 0$
for some $n$.
 It was shown in \cite{deMedeiros:2002ge} that the natural  equation of motion
 \be \t \,F=0 \qquad  \text{for} \qquad  D\ge p+q+2 \ ,
 \ee
is non trivial for $D\ge p+q+2$, but for $D<p+q+2$ that $ \t^n \,F = 0$
 is a non-trivial field equation
  for $n = p+q+3 - D$, and so we will take
 \be \t^{ p+q+3 - D}\,F = 0 \qquad
 \text{for} \qquad  D<p+q+2 \ .
\ee

 A $[p,q]$-Young tableau can be dualized   on one of the two columns, or
on both, so that three duals of the field strength can be defined:
\be S \equiv \ast \,F \in X^{D-p-1,q+1} \ ,
\hspace{.5cm} \tilde{S} \equiv \tilde{\ast} \,F \in X^{p+1,D-q-1} \ ,
\hspace{.5cm} G \equiv \tilde{\ast}\,\ast \,F \in X^{D-p-1,D-q-1} \ ,
\ee
and the  algebraic and differential
identities and equations of motion for $F$ give
analogous properties for $S$, $\tilde{S}$ and $G$. In particular,
the equations of motion are
\footnote{The last equation follows from the result
  $\t^n \,T = 0 \Rightarrow (\t^{D-p-q+n} \,\ast \,\tilde{\ast})\,T = 0 \ ,$
valid for a general [p,q]-form $T$, applied to the case of the [p+1,q+1]-form $F$. Here
$n$ is the exponent such that $\t^n \,F=0$ is non trivial. So, if
$D\geq p+q+2$ then $n=1$, and the e.o.m. for the dual tensor $G$ reduces to
$\t^{D-p-q-1}\,G = 0$ \cite{deMedeiros:2002ge} \ .}:
\be
\t \,S=0 \ , \hspace{1.5cm}
\t^{1+p-q}\,\tilde{S} = 0 \ , \hspace{1.5cm} \t^{D-p-q-2+n}\,G = 0 \ .
\ee
For example, gravity in $D=3$ with $p=q=1$ has dual formulations in terms of
a $[1,1]$ field strength $G$ for a $[0,0]$ form or scalar field $C$ with field equation
$\t \,G=0$ giving the usual scalar  field equation $\partial _\m\,\partial ^\m \,C=0$, or to a
$[2,1]$ field strength $S$ for a $[1,0]$ or vector gauge field $D_\m$, with the usual Maxwell equation
$\t \,S=0$. Then this $D=3$ gravity theory is dual to a scalar  field and to a
vector field, and all describe one physical degree of freedom.

\section{Multi-Forms}

 The previous discussion generalizes to the case of fields in
 arbitrary massless representations
of $SO(D-1,1)$, including higher-spin gauge fields described by
mixed-symmetry Young tableaux. As for the case of forms and bi-forms, the starting point is the definition of a larger
environment, the space of \emph{multi-forms}, in which a series of
useful operations are easily defined. Then, by suitable Young
projections, one can discuss the cases of irreducible gauge fields
and their duality properties. In the following we shall confine
ourselves to describing the main steps of the construction; further
details are given in \cite{deMedeiros:2002ge},\ \cite{deMedeiros:2003dc}.

A multi-form of order $N$  is characterised by a set of $N$ integers
$(p_1,p_ 2,...,p_N)$
and is a tensor of rank  $\sum p_i$
 whose components
\be
T_{\m_1^1 \dots \m_{p_1}^1 \dots \m_1^N \dots \m_{p_N}^N}=T_{[\m_1^1 \dots \m_{p_1}^1] \dots [\m_1^N \dots \m_{p_N}^N]} \ ,
\ee
are totally antisymmetrized within each of $N$ groups of  $p_i$ indices,
with no  other symmetry \emph{a priori} between indices
belonging to different sets.
It is an element of
$X^{p_1\dots p_N} \equiv \L^{p_1} \otimes \dots \otimes~\L^{p_N}$,
the $GL(D,\mathbb{R})$-reducible  $N$-fold tensor product space
of $p_i$-forms on $\mathbb{R}^D$.
%A  \emph{multi-form} $T$ is a generic element of $X^{p_1\dots p_N}$.
The operations and the properties
introduced in Section $4$ generalize easily to multi-forms. For an
extensive treatment see again \cite{deMedeiros:2002ge}; here we
restrict our attention to the operations previously discussed.

One can define an exterior derivative acting on the $i-th$ set of
indices,
\be
d^{(i)}: X^{p_1\dots p_i \dots  p_N} \ra X^{p_1 \dots
p_{i+1} \dots p_N} \ ,
\ee
generalizing the properties of $d$ and
$\tilde{d}$; summing over the $d^{(i)}$'s one can then define the
\emph{total derivative}
\be
{\cal{D}} \equiv \sum_{i=1}^{N} d^{(i)}\, ,
\ee
such that
\be
{\cal{D}}^{N+1} = 0 \ .
\ee
Similarly,
for representations of $SO(D-1,1)$  or $SO(D )$
one can define $N$ Hodge-duals:
\be \ast^{(i)} : X^{p_1,\dots, p_i, \dots, p_N} \ra X^{p_1, \dots,
D-p_{i}, \dots, p_N} \ , \ee each acting in the usual fashion on
the $i$-th form, and so commuting with any $\ast^{(j \neq i)}$.

 The operators $\t$, $\s$, $\tilde{\s}$ and $t$ generalize
to a set of operators, each acting on a specific pair of indices;
the \emph{trace} operators
\be
\t^{(ij)} : X^{p_1, \dots,p_i,
\dots, p_j, \dots, p_N} \ra X^{p_1, \dots,p_{i-1}, \dots, p_{j-1},
\dots, p_N} \ ,
\ee
are defined as traces over the $i$-th and the
$j$-th set; the \emph{dual-traces} are
\be
\begin{split}
\s^{(ij)} & \equiv (-1)^{1+D(p_i+1)} \ast^{(i)} \t^{(ij)} \ast^{(i)} :
X^{p_1, \dots,p_i, \dots, p_j, \dots, p_N} \ra X^{p_1, \dots, p_{i+1}, \dots,p_{j-1}, \dots  p_N} \ , \\
\tilde{\s}^{(ij)} & \equiv (-1)^{1+D(p_j+1)} \ast^{(j)}  \t^{(ij)}
\ast^{(j)} : X^{p_1, \dots,p_i, \dots, p_j, \dots, p_N} \ra
X^{p_1, \dots, p_{i-1}, \dots,p_{j+1}, \dots  p_N} \ ,
\end{split}
\ee
while the \emph{transpositions} $t^{(ij)}$  generalize the action
of the $t$ operator to
exchanges between the subspaces $\L^{p_i}$ and $\L^{p_j}$ in
$X^{p_1, \dots,p_i, \dots, p_j, \dots, p_N}$:
\be
t^{(ij)} :
X^{p_1, \dots,p_i, \dots, p_j, \dots, p_N} \ra X^{p_1, \dots,p_j,
\dots, p_i, \dots, p_N}.
\ee
 The Young symmetrizer ${\cal{Y}}_{[p_1,
\dots, p_N]}$ projects a multi-form of type $(p_1,
\dots, p_N)$ onto the irreducible representation associated with a Young tableau of type
 $[p_1,
\dots, p_N]$.

\section{Multi-Form Gauge Theories}

With the machinery of the last section, one can naturally extend the construction
of gauge theories for  general tensor gauge  fields.
The starting point is a multi-form gauge field
of type $(p_1,
\dots, p_N)$ with gauge transformation
\be
\d \,T = \sum_{i=1}^N d^{(i)} \,\a_{(i)}^{p_1,
\dots, p_{i-1},\dots, p_N} \ .
\ee
The
restriction to irreducible representations of $GL(D,\mathbb{R})$
can be implemented using the
Young symmetrizer ${\cal{Y}}_{[p_1,
\dots, p_N]}$
projecting onto the representation characterised by a Young tableau with $N$
columns of length $p_1,p_2,...,p_N$ (these are conventionally arranged in order of decreasing length, but this is not essential here).
Then this projects a multi-form $T$ onto
\be
\widehat T={\cal{Y}}_{[p_1,
\dots, p_N]}\,T
\ee
which satisfies
 the constraints \be
\begin{split}
\s^{ij} \,\widehat T & = 0 \ \qquad \text{if} \ \ \ ~p_i \ge p_j \\
t^{ij} \,\widehat T  & = \widehat T \ \qquad \text{if} \ \ \ ~p_i = p_j
\end{split}
\ee
 The field strength is a multi-form in the irreducible representation of type $ [p_1+1,\dots,p_N~+~1]  $ defined
as\footnote{More generally, one can define a
set of connections $\G_{S_k} \equiv (\prod_{i \not\in S_k} d^{(i)}) \,\widehat T$, corresponding to each subset $S_k =\{i_1, \dots, i_k \} \subseteq \{1, \dots , N\}$. These  are
gauge-dependent w.r.t. transformations involving parameters
$\a^j$\ , $j \in S_k$\ , while are invariant under transformations with parameters
$\a^i \ , i\not\in S_k\ .$ For a given $k$ there are in general $\fr{N!}{(N-k)!k!}$ inequivalent
possible $\G_{S_k}$; in particular, the totally gauge-invariant field strenght $F$
can be regarded both as the top of this hierarchy of connections (the one with $k=0$), or as a
direct function of the connection $\G_{S_k}$, being $F= (\prod_{i \in S_k} d^{(i)})
\,\G_{S_k}$
\cite{deMedeiros:2003dc}\ .}
\be
F \equiv \prod_{i=1}^N d^{(i)} \,\widehat{T} =
\fr{1}{N} {\cal{D}}^N \,\widehat{T}\ ,
\ee
and is invariant under the Young projection of the gauge
transformation for $T$
\be
\d \,\widehat T ={\cal{Y}}_{[p_1,
\dots, p_N]}
 \sum_{i=1}^N d^{(i)} \,\a_{(i)}^{p_1,
\dots, p_{i-1},\dots, p_N} \ .
\ee
By construction, the field
strength satisfies the generalized Bianchi identities
\ba
d^{(i)} \,F = 0 \ , \\
\s^{(ij)} \,F = 0 \ .
\ea
The simplest covariant
local field equations are those  proposed in \cite{deMedeiros:2002ge} and  these in general
involve more than two derivatives.
For $N$ even, a suitable field equation is
\be \label{eomeven}
\sum_{\substack{(i_1, \dots, i_N)  \\
      (  permutations)}} \t^{(i_1i_2)} \dots \t^{(i_{N-1} i_N)} \,F = 0 \ ,
\ee
where the sum is over all permutations of the elements of the set
$\{1, \dots N \}$.
For $N$ odd, one first needs
to define  $ \pr \,F$,  which is  the derivative $ \pr _\m \,F$ of $F$
regarded as a rank $N+1$ multi-form of type $[p_1, \dots, p_N,1]$.
Then the equation of motion is:
\be \label{eomodd}
\sum_{\substack{(i_1, \dots, i_N)  \\
      (  permutations)}} \t^{(i_1i_2)} \dots \t^{(i_{N-2} i_{N-1})} \t^{(i_{N} N+1)} \,\pr \,F = 0 \ .
\ee Here the sum is over \emph{the same} set of
permutations of the elements of the set
$\{1, \dots N \}$ as in the even case,
so that the extra index is left out.
These are the field equations for large enough space-time dimension $D$; as for the case of bi-forms,
for low dimensions one needs to act with further traces.

 These field equations involving multiple traces of a higher-derivative
tensor  are necessarily of higher order in derivatives if $N>2$. This is   unavoidable if
the field equation for a higher-spin field is to be written in terms of invariant  curvatures.
In physical gauge, these field equations become
$\Box^a A=0$ where $A$ is the gauge potential in physical gauge,
$\Box$\ is the $D$-dimensional
d'Alembertian operator
and $a=N/2$ if $N$ is even and
$a=(N+1)/2$ if $N$ is odd. The full covariant field equation is of order $2a$ in derivatives.
 In order to get a second order
equation,   following \cite{Francia:2002aa,Francia:2002pt},
one can act on these covariant field equations with  $\Box^{1-a}$
to obtain equations that reduce to the second order equation
$\Box A=0$   in physical gauge.
In the even case the equation (\ref{eomeven}) is of order $N$ in derivatives,
and so it is possible to write a second-order field equation  dividing by $\Box^{\fr{N}{2}-1}$.
Similarly, for $N$ odd, it is necessary to divide by $\Box^{\fr{N+1}{2} -1}$. In this way, one can write
second-order, non-local  field equations \cite{deMedeiros:2003dc}:
\be
\begin{split} \label{eomnonloc}
 {\cal{G}}_{\text{even}} & \equiv  \sum_{\substack{(i_1, \dots, i_N)  \\
      (  permutations)}} \t^{(i_1i_2)} \dots \t^{(i_{N-1} i_N)} \fr{1}{\Box^{\fr{N}{2}-1}}\,F = 0 \ ,\\
 {\cal{G}}_{\text{odd}} & \equiv  \sum_{\substack{(i_1, \dots, i_N)  \\
      (  permutations)}} \t^{(i_1i_2)} \dots  \t^{(i_{N} N+1)}\fr{1}{\Box^{\fr{N+1}{2}-1}} \,\pr \,F = 0 \ .
\end{split}
\ee

These then are the covariant field equations for general representations for high enough $D$ (for low $D$,
the appropriate field equations require further traces \cite{deMedeiros:2003dc}, as we saw earlier for the case of bi-forms.)
These are non-local, but after fixing a suitable gauge, they become local.
On fully fixing the gauge symmetry to go to light-cone gauge, the field equations reduce to
the free equation $\Box A=0$, while partially fixing the gauge gives a Fronsdal-like local
covariant field equation with constraints on the traces of the gauge field and parameters of the surviving gauge symmetries.
It would be interesting to understand if the non-locality of the full geometric field equation has
any physical consequences, or is purely a gauge artifact. As in the Fronsdal case,
only physical polarizations are propagating \cite{Francia:2002pt,deMedeiros:2003dc,Bekaert:2003az}.

It is worth noting that these equations are not unique. As was observed in \cite{Francia:2002aa}, and analysed in detail for the case $s=3$ in the totally symmetric representation, one can write other second-order equations, with
higher degree of non locality, by combining the least singular non-local equation
with its traces and divergences. The systematics of this phenomenon was described in \cite{deMedeiros:2003dc},
where it was shown in the general case how to generate other field equations starting from
(\ref{eomnonloc}). The idea is to define a new tensor $F^{(m)} \equiv \pr^m F$,
by taking $m$ partial derivatives of the field strenght $F$, take a suitable number of traces of the
order $N+m$ tensor $\pr^m F$, and divide for the
right power of the D'Alembertian operator.
One can then take linear combinations of these  equations with the original equations
(\ref{eomnonloc}).

Given a   field strength $F$ of type  $ [p_1+1,p_2+1, \dots, p_N+1]  $, one can choose any set
of columns of the Young tableau and dualise on them to obtain a dual field strength.
The field equations and Bianchi identities for
$F$ then give the field equations and Bianchi identities for the dual field strength,
and the new Bianchi identities imply that the dual field strength can be solved for in terms of a dual potential.
There are then many dual descriptions of the same free higher-spin gauge  theory.

%\section{Summary and outlook}
\vskip 24pt \noindent{\large \bf Acknowledgements}

D.F. would like to thank the Organizers for giving him the
opportunity to participate in the Workshop and to collect and
order these lecture notes. The work of D.F. was supported in part
by I.N.F.N., by the MIUR-COFIN contract 2003-023852, by the INTAS
contract 03-51-6346 and by the NATO grant PST.CLG.978785.
We would like to thank Paul de Medeiros for comments on the manuscript.

%%%%%%%%%%%%%%%%%%%%%%%%%%%%%%%%%%%%%%%%%%%%%%%%%%%%%%%%%%%%%%%%%%%%%%%%%
\end{document}